# Criteria for objects suitable for reconstruction from holograms and diffraction patterns

by Tatiana Latychevskaia

tatiana@physik.uzh.ch, tatiana.latychevskaia@psi.ch

## Abstract

In this study, quantitative criteria for reconstruction of objects from their hologram and diffraction patterns, and in particular for the phase objects in digital holography, are derived.

The criteria that allow distinguishing the hologram and diffraction pattern are outlined.

Gabor derived his criterion for objects suitable for holography based on the condition that the background in the reconstructed object's distribution should be nearly flat so that its intensity contrast does not exceed 0.05. According to Gabor, an opaque object is suitable for holographic reconstruction if it occupies no more than 1% of the imaged area, and a phase-shifting object cannot be reconstructed in principle. We revisit these criteria and show that both amplitude-only and phase-only objects can be reconstructed when the object occupies less than 1% of the total illuminated area.

In addition, a simplified derivation of the criteria is provided that is based on Parseval's theorem. It is shown that for objects (including amplitude-only and phase-only) reconstructed from their holograms and the twin image treated as noise, the signal-to-noise ratio of 10 or higher can be achieved provided the object occupies less than 0.5% of the total illuminated area.

When a hologram is reconstructed by applying iterative algorithms, the requirement for the object size is much more generous and identical to that applied in coherent diffraction imaging: any type of object (amplitude-only, phase-only, or amplitude-and-phase mixed properties) is suitable for holography when the object's size in each dimension is less than half of the probed region's extent (or the field of view).

## Content

# Main text

# 1. Introduction

Coherent lensless imaging techniques produce images of the sample without using a lens system but by using an analysis of the interference pattern created by the wave scattered or diffracted by the sample. The most common coherent lensless imaging techniques are summarized in Fig. 1: Gabor (or in-line) holography with spherical wave (Fig. 1a), in-line holography with plane wave (Fig. 1b), off-axis holography (Fig. 1c), and coherent diffraction imaging (CDI) (Fig. 1d). In holographic schemes (Fig. 1a – c) [1, 2] (Gabor1948Nature, Gabor1949ProcRoySocLondA), the presence of the reference wave allows direct capture of the phase of the object wave. The sample structure is then reconstructed simply by wavefront backward propagation. In coherent diffraction imaging (CDI) (Fig. 1d), the phase information is lost and typically reconstructed by applying iterative phase retrieval (IPR) algorithms [3-5] (Miao1999Nature, Fienup1978OptLett, Fienup1982JOSA). IPR algorithms have also been demonstrated for reconstructing objects from their holograms, which allows to eliminate the twin images and achieve quantitatively accurate reconstruction of the absorption and phase distributions of the imaged objects [6, 7] (Latychevskaia2007PRL, Latychevskaia2019JOSA).

The condition for an object to be suitable for holography was outlined by Gabor [2] (Gabor1949ProcRoySocLondA): an opaque object should not occupy less than 1% of the imaged region. A similar condition is applied in CDI: an object can be reconstructed from its diffraction pattern only if the object's size in each dimension is less than half of the probed region or field of view [8] (Miao2000ActaCrystallogrSectA). In this study, we address the quantitative difference between hologram and diffraction pattern, outline the quantitative criteria for amplitude-only and phase-only objects to be reconstructed from their holograms, and discuss the related reconstruction procedures.

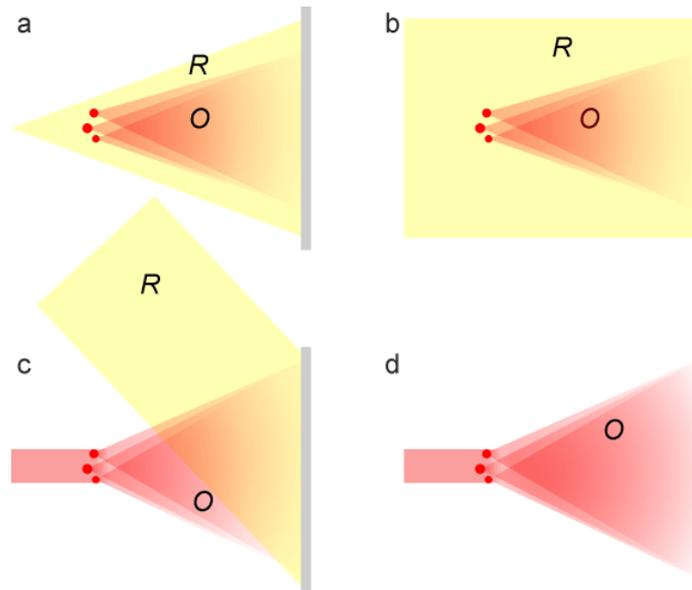

Fig. 1. Overview of most common coherent lensless imaging schemes: (a) Gabor or in-line holography with spherical wave. (b) In-line holography with plane wave. (c) Off-axis holography. (d) Coherent diffraction imaging (CDI). *R* – reference wave (shown in yellow), *O* – object wave (shown in red).

## 2. Difference between hologram and diffraction pattern

In the case of in-line (or Gabor type) holography and CDI, a clear separation between when the recorded interference pattern constitutes a "hologram" and when it is a "diffraction pattern" is important because, firstly, the two types of images require different reconstruction methods, and secondly, different types and limits of information can be reconstructed. The difference between a hologram and a diffraction pattern is illustrated in Fig. 2. For clarity, we will refer to the entire imaged (or illuminated) region as "sample distribution", and the imaged object in this region we will name as "object". In in-line (Gabor) holography, the object occupies a small fraction of the illuminated area, so that a large part of the wave does not interact with the object and provides the reference wave. The two waves – the object wave, which interacted with the object, and the reference wave, which did not interact with the object – produce an interference pattern or hologram. The following points are fulfilled for an in-line hologram:

(i) A hologram is an interference pattern created by interference between the object wave (the wave that interacted with the object) and the reference wave (the wave that did not interact with the object) (Fig. 2b).

(ii) A Hologram is typically acquired in the Fresnel diffraction regime [9] (Goodman2004Introduction) which is described by the condition

$$z \gg \left\{ \frac{\pi}{4\lambda} \left[ (X-x)^2 + (Y-y)^2 \right]^2_{\max} \right\}^{1/3}, \quad (1)$$

where $(x, y)$ are the coordinates in the sample plane, $(X, Y)$ are the coordinates in the detector plane, $z$ is the distance between the sample and detector planes, and $\lambda$ is the wavelength. Equation (1) describes the case when the object is illuminated by a plane wave. When the object is illuminated by a spherical wave, the diffracted wavefront distribution is similar to the plane wave diffracted by the object in the Fresnel diffraction regime; the parameters can be re-calculated using the relations provided in ref. [10] (Latychevskaia2015AO).

(iii) The object distribution can be recognized in its hologram (Fig. 2b).

(iv) A reconstruction of the object can be obtained by back propagation of the wavefront from the hologram plane to the sample plane.

In CDI, a diffraction pattern is a recorded intensity distribution of the wave scattered or diffracted by an object without the presence of a reference wave, so that the phase information is completely lost (Fig. 1(d)). Unlike in holography, the sample distribution cannot be recovered by simple back propagation of the wavefront from the detector plane to the sample plane (that is, by calculating the inverse Fourier transform of the diffraction pattern). The distribution obtained by backward propagation of the wavefront will not resemble the original sample distribution. The sample distribution can be reconstructed only by applying IPR algorithms. The most popular IPR algorithms are the error-reduction (ER) algorithm, the hybrid input output (HIO) algorithm, and their variations [5, 11, 12] (Fienup1982JOSA, Marchesini2003PRL, Latychevskaia2018AO).

In CDI, the following points are fulfilled for a diffraction pattern:

(i) In diffraction pattern, the interference only between the scattered (diffracted) wave with itself is recorded, without a reference wave (Fig. 2c).

(ii) Diffraction pattern is typically acquired in the Fraunhofer diffraction regime [9] (Goodman2004Introduction),

$$z \gg \frac{\pi}{\lambda} \left( x^2 + y^2 \right)_{\max}, \quad (2)$$

which, for a plane incident wave, means it is acquired in far field or at a large distance from the sample.

(iii) The sample distribution cannot be recognized in its diffraction pattern (Fig. 2c).

(iv) Reconstruction of the sample or object can be obtained only by applying IPR methods.

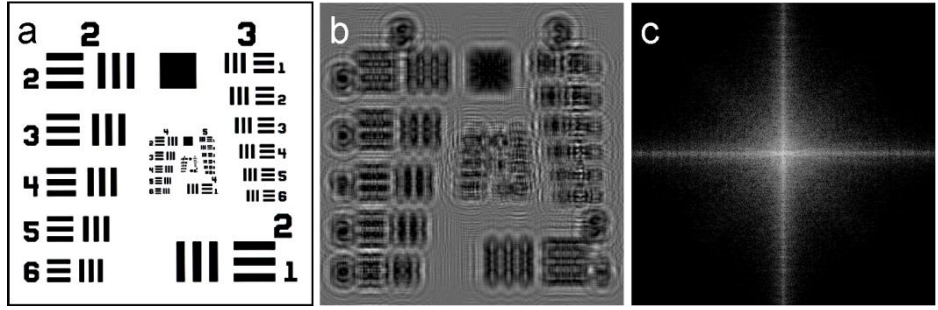

Fig. 2. Illustration to the difference between hologram and diffraction pattern. (a) Sample distribution (amplitude-only), (b) its in-line hologram, and (c) its diffraction pattern.

## 3. CDI

In CDI, an object can be reconstructed from its diffraction pattern, provided that the object's size in each dimension is less than half of the probed region's extent [3, 8, 13] (Miao1999Nature,Miao2000ActaCrystA,Miao2000ActaCrystD). This criterion is derived as follows. The intensity of the diffraction pattern is given by the Fourier transform of the sample distribution:

$$I(p,q) = |O(p,q)|^2 \propto \left| \sum_{m,n=0}^{N-1} o(m,n) \exp\left[ -\frac{2\pi i}{N}(mp+nq) \right] \right|^2, \quad (3)$$

where $o(m,n)$ is the sample distribution, $O(p,q)$ is its Fourier transform, $(m,n)$ and $(p,q)$ are the coordinates (in pixels) in the sample and detector plane, respectively, $m,n,p,q = 0...N-1$ are the running pixel indices. In Eq. (3), the phase factors that account for centering the sample and detector-plane distributions at the optical axis are omitted [10] (Latychevskaia2015AO). Both, the sample and the intensity in the detector plane distributions, are sampled with $N \times N$ pixels, and with this, Eq. (3) represents a set of $N^2$ equations. The object is sampled with $N_0 \times N_0$ pixels (Fig. 3). This gives $N_0^2$ unknowns for a real-valued object; since the diffraction pattern is symmetrical for a real-valued object, there are only $N^2/2$ equations in Eq. (3) in this case. For a complex-valued object, there are $2N_0^2$ unknowns, and there are $N^2$ equations in Eq. (3). For a system of equations to have a solution, the number of unknowns should be less than the number of equations. This leads to the condition: $N_0 < N/\sqrt{2}$, for both real- and complex-valued objects. By introducing an oversampling parameter $\sigma = N/N_0$, the condition can be rewritten as $\sigma > 2$ (approximately). This condition should be fulfilled in each dimension (in the x- and y-directions). In practice, the parameters of the experimental setup are set to provide as large an oversampling ratio as possible to ensure faster convergence of the IPR routine [12] (Latychevskaia2018AO).

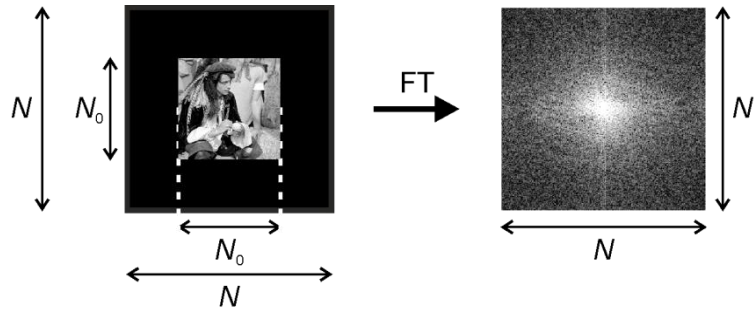

Fig. 3. Illustration to the oversampling condition in coherent diffraction imaging. Left: object sampled with $N_0 \times N_0$ pixels, the sample area is sampled with $N \times N$ pixels. Right: diffraction pattern, sampled with $N \times N$ pixels. FT means Fourier transform.

## 4. Holography

In holography, the criterion for objects to be suitable for holographic imaging were outlined by Gabor, based on the condition that the non-object terms present in the reconstructed signal should create no more than 5% intensity contrast relatively to the correctly reconstructed object term. In this section, we reproduce in detail and revise Gabor's derivation of the criterion for an object suitable for holographic reconstruction [2] (Gabor1949ProcRoySocLondA).

### 4.1 Gabor's criterion for object suitable for holography

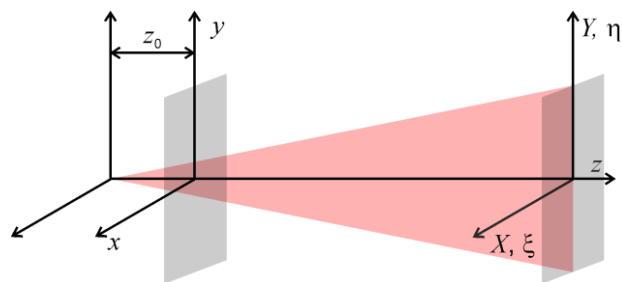

Fig. 4. Illustration to the symbols used in hologram formation theory.

In Gabor geometry, the sample is illuminated by a spherical wave; the schematics and symbols are illustrated in Fig. 4. The wavefronts in the sample and detector planes constitute a pair of the following transformations (Eqs. (14) and (15) in ref. [2] (Gabor1949)):

$$\tau(\xi,\eta) = \frac{1}{i\mu} e^{i\pi\mu\rho^2} \iint t(x,y) e^{-\pi r^2/i\mu} e^{-2\pi i(x\xi+y\eta)} \mathrm{d}x\mathrm{d}y, \tag{4}$$

$$t(x,y) = i\mu e^{\pi r^2/i\mu} \iint \tau(\xi,\eta) e^{-i\pi\mu\rho^2} e^{+2\pi i(x\xi+y\eta)} \mathrm{d}\xi\mathrm{d}\eta, \tag{5}$$

where $r^2 = x^2 + y^2$, $\rho^2 = \xi^2 + \eta^2$, $(x,y)$ are the coordinates in the sample plane, $(\xi,\eta)$ are the "Fourier" coordinates in the far field, their connection with the coordinates in the detector plane $(X,Y)$ at the distance $z = L$ is given by

$$X = \lambda L \xi / \left[1 - \lambda^2 \left(\xi^2 + \eta^2\right)\right]^{1/2}, \quad Y = \lambda L \eta / \left[1 - \lambda^2 \left(\xi^2 + \eta^2\right)\right]^{1/2}, \tag{6}$$

$\mu$ is given by

$$\mu = \lambda z_0, \tag{7}$$

and $z_0$ is the source-to-sample distance. From Eq. (4) – (5), it follows that for $t(x,y) = 1$, its counterpart $\tau(\xi,\eta) = 1$.

Next, as the smallest object, Gabor considers a "probability spot" – an object with a Gaussian distribution. Note that here one cannot consider a delta-like function as the smallest object because, in order to be able to separate the object and reference waves, the object must have a finite size. Another choice would be a rectangular function that would correspond to a single pixel (in our days of digital holography), but it is also not a good choice because it would result in integrals that cannot be solved analytically. Gaussian distribution, on the other hand, is a very practical distribution because its Fourier transform is also a Gaussian distribution, and thus, the expressions for hologram and reconstructed object distributions can be derived analytically.

According to Gabor, the sample distribution containing a single "probability spot" located at $(x_0, y_0)$ is described by the transmission function:

$$t(x,y) = 1 - A\exp\left(-\frac{r'^2}{a^2}\right), \tag{8}$$

where $A$ is a complex-valued constant, $a$ is a parameter that describes the spread of the Gaussian function (note that it is different by factor $\sqrt{2}$ from the standard deviation that is used in the conventional definition of Gaussian distribution), $r'^2 = (x-x_0)^2 + (y-y_0)^2$ is the distance between the center of the "probability spot" and coordinate $(x,y)$ in the sample plane.

Equation 8, in general, describes a complex-valued transmission function that can be re-written as $t(x,y) = 1 + o(x,y)$. Here, the first term ("1") describes the part of the wave that goes unchanged through the sample plane and creates a uniform intensity distribution in the detector plane, and the second term – $o(x,y)$ – describes any deviation from the uniform wavefront of "1" due to the presence of the object. By substituting Eq. (8) into Eq. (4) (or using the transformation pair in Appendix 1), one obtains the analytical expression for the distribution of the wavefront in the detector plane (Eq. (21) in ref. [2] (Gabor1949ProcRoySocLondA)):

$$\tau(\xi,\eta) = 1 + \frac{i\varepsilon A}{1-i\varepsilon} \exp\left(\frac{i\pi\mu\rho'^2}{1-i\varepsilon}\right), \tag{9}$$

where dimensionless parameter $\varepsilon$ is introduced as

$$\varepsilon = \frac{a^2}{\mu} = \frac{a^2}{\lambda z_0}, \tag{10}$$

and $\rho'^2 = (\xi-\xi_0)^2 + (\eta-\eta_0)^2$ is the distance between the centre of the shadow of the "probability spot" located at $(\xi_0,\eta_0)$ and a point $(\xi,\eta)$ in the detector plane. The hologram distribution is given by

$$\tau_s(\xi,\eta) = |\tau(\xi,\eta)|^2 = \left|1 + \frac{i\varepsilon A}{1-i\varepsilon} \exp\left(\frac{i\pi\mu\rho'^2}{1-i\varepsilon}\right)\right|^2. \tag{11}$$

By neglecting small $\varepsilon^2$ terms, the hologram's distribution can be re-written as (Eq. (22) in ref. [2] (Gabor1949ProcRoySocLondA)):

$$\begin{aligned}\tau_s(\xi,\eta) &\approx \left|1 + i\varepsilon A \exp\left[i\pi\mu(1+i\varepsilon)\rho'^2\right]\right|^2 \\ &\approx 1 + i\varepsilon A \exp(i\pi\mu\rho'^2)\exp(-\pi\mu\varepsilon\rho'^2) - i\varepsilon A^* \exp(-i\pi\mu\rho'^2)\exp(-\pi\mu\varepsilon\rho'^2) = \\ &= 1 + i\varepsilon A \exp(i\pi\mu\rho'^2)\exp\left[-\pi(a\rho')^2\right] - i\varepsilon A^* \exp(-i\pi\mu\rho'^2)\exp\left[-\pi(a\rho')^2\right],\end{aligned} \tag{12}$$

where we took into account that $\mu\varepsilon = a^2$ from Eq. (10). The reconstructed distribution is given by substituting of Eq. (12) into Eq. (5) or using the transformation pair in Appendix 2 (Eq. (22.1) in ref. [2] (Gabor1949ProcRoySocLondA)):

$$\begin{aligned}t_r(x,y) &= 1 - A\exp\left[-\pi\left(\frac{r'}{a}\right)^2\right] - \frac{i\varepsilon A^*}{2+i\varepsilon} \exp\left[-\frac{\pi(\varepsilon+2i)}{\mu(4+\varepsilon^2)}r'^2\right] + \\ &+ \frac{\varepsilon^2 AA^*}{1+\varepsilon^2-2i\varepsilon} \exp\left\{-\frac{2\pi\varepsilon(1+\varepsilon^2+2i\varepsilon)}{\mu\left[(1+\varepsilon^2)^2+4\varepsilon^2\right]}r'^2\right\}.\end{aligned} \tag{13}$$

A simulated hologram of an amplitude-only probability spot, described by Eq. (8) with $A=1$: $t(x,y) = 1 - \exp\left(-\frac{r^2}{a^2}\right)$, together with the reconstructed distribution, is shown in Fig. 5. A simulated hologram of a phase-only probability spot, described by Eq. (8) as $t(x,y) = \exp[i\varphi(x,y)]$ where $\varphi(x,y) = \varphi_0 \exp\left(-\dfrac{r^2}{a^2}\right)$, together with the reconstructed distribution, is shown in Fig. 6; the case of phase-shifting objects is discussed later in more detail. The simulations were done using the following parameters: wavelength = 500 nm, object area size 2 × 2 mm$^2$, source-to-sample distance $z_0$ = 40mm, source-to-detector distance $z$ = 0.2 m, $a$ = 0.05 mm, the distributions were sampled with 200 × 200 pixels, only 100 × 100 pixels regions are shown.

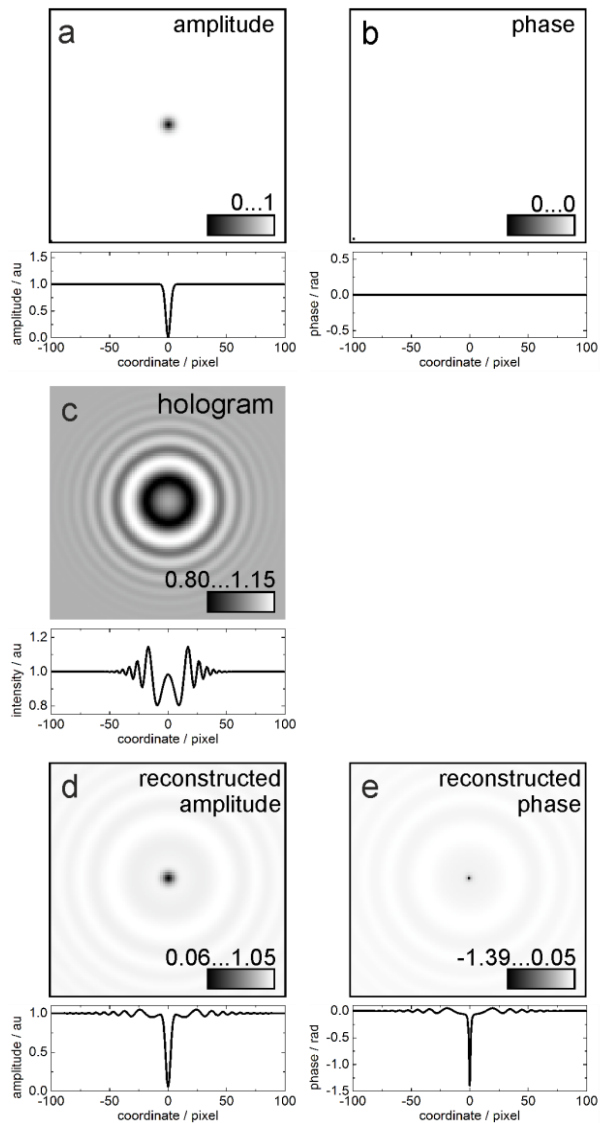

Fig. 5. Amplitude-only "probability spot", its hologram and reconstruction. (a) amplitude and (b) phase distributions of an amplitude-only "probability spot". The transmission function is given by $t(x,y) = \exp(-r^2/a^2)$. (c) The corresponding hologram, and reconstructed (d) amplitude and (e) phase distributions, respectively. The plots below

the two-dimensional distributions show the horizontal profiles through the middle of the two-dimensional distributions.

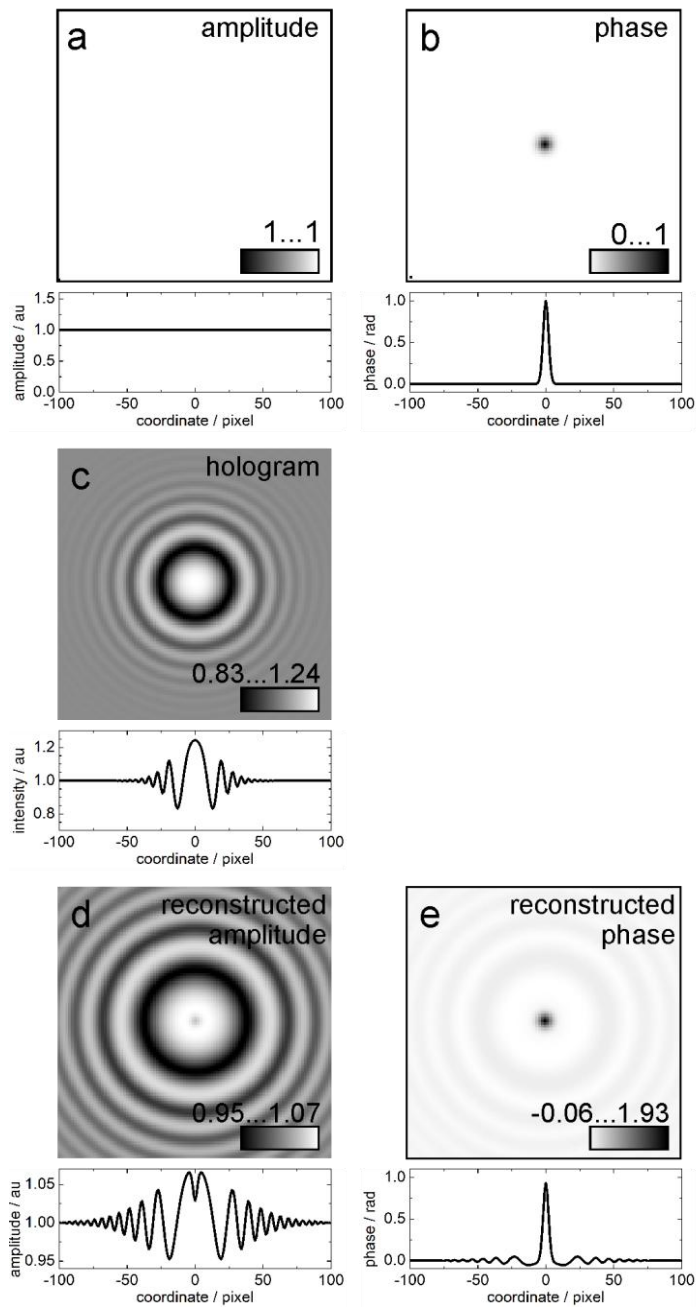

Fig. 6. Phase-only "probability spot", its hologram and reconstruction. (a) amplitude and (b) phase distributions of a phase-only "probability spot". The transmission function is given by $t(x,y) = \exp[i\varphi(x,y)]$ where $\varphi(x,y) = \varphi_0 \exp(-r^2/a^2)$, $\varphi_0 = 1$ rad. (c) The corresponding hologram, and reconstructed (d) amplitude and (e) phase distributions, respectively. The contrast of the two-dimensional phase distributions in (b) and (e) is inverted for showing a printer-friendly image. The plots below the two-

dimensional distributions show the horizontal profiles through the middle of the two-dimensional distributions.

The resolution of the reconstructed object is defined (among other factors) by the hologram's effective size, that is, by the extent of the actual interference pattern, which can be limited either by the detector size or noise. Gabor assumes that the hologram is masked by a function with a Gaussian distribution:

$$\exp\left[-\pi(b\rho)^2\right]. \tag{14}$$

In digital holography, the hologram distribution can be multiplied by a cosine-like mask function [10] (Latychevskaia2015AO)

$$c(\rho) = \begin{cases} 1, & \rho \leq \rho_m \\ \cos^2\left[\dfrac{\pi}{2w_m}(\rho - \rho_m)\right], & \rho_m < \rho < \rho_m + w_m \\ 0, & \rho \geq \rho_m + w_m \end{cases} \tag{15}$$

before the reconstruction procedure to avoid signal wrapping effects at the hologram's edges that arise during application of fast Fourier transforms (FFTs); here $\rho_m$ is the radius of the mask and $w_m$ is the width of the mask. The mask given by Eq. (15) has a similar effect as the Gaussian function in Eq. (14), but its parameters can be adjusted such that the smoothing is applied only at the hologram's edges, keeping the values in the hologram's centre unchanged. For the sake of following Gabor's derivation, we keep the masking function in the form of Eq. (14).

The transmission function of the hologram is then given by multiplying Eq. (12) with Eq. (14) (Eq. (24) in ref. [2] (Gabor1949ProcRoySocLondA)):

$$\begin{aligned}\tau_s(\xi,\eta) &\approx \exp\left[-\pi(b\rho)^2\right] \times \\ &\times \left\{1 + i\varepsilon A \exp(i\pi\mu\rho'^2)\exp\left[-\pi(a\rho')^2\right] - i\varepsilon A^* \exp(-i\pi\mu\rho'^2)\exp\left[-\pi(a\rho')^2\right]\right\},\end{aligned} \tag{16}$$

where we neglected small $\varepsilon^2$ terms. The reconstructed distribution is obtained by substitution of Eq. (16) into Eq. (5) or using the transformation pair in Appendix 2 (Eq. (25) in ref. [2] (Gabor1949ProcRoySocLondA)):

$$\begin{aligned}t_r(x,y) &\approx (1+i\sigma)\exp\left[-\frac{\pi\sigma}{\mu}(1+i\sigma)r^2\right] - \frac{A}{1+(b/a)^2}\exp\left\{-\frac{\pi}{a^2+b^2}\left[r'^2 + \varepsilon\sigma r_0^2 + i(\varepsilon-\sigma)r_0^2\right]\right\} - \\ &-\frac{1}{2}i\varepsilon A^*\exp\left(-\frac{\pi i r'^2}{2\mu} - \frac{\pi}{4\mu}\left\{\varepsilon\left[(x-x_0)^2+(y-y_0)^2\right] + \sigma\left[(x+x_0)^2+(y+y_0)^2\right]\right\}\right),\end{aligned}$$
(17)

where $r_0^2 = x_0^2 + y_0^2$, and the dimensionless parameter $\sigma$ is given by:

$$\sigma = \frac{b^2}{\mu} = \frac{b^2}{\lambda z_0}. \tag{18}$$

The first term in Eq. (17) is the background term; it is the geometrical shadow of the mask, projected on the object plane, and its amplitude follows a Gaussian distribution.

The second term in Eq. (17) is the reconstructed object – the Gaussian distribution, but with the standard deviation proportional to $(a^2 + b^2)$ instead of $a^2$. Thus, $b$ has the meaning of the resolution limit. It should be noted that Gabor omits the term with $(\varepsilon - \sigma)$; this term disappears in the following derivation where it is set $\varepsilon = \sigma$.

The third term in Eq. (17) is the twin image or error term (Eq. (25.1) in ref. [2] (Gabor1949ProcRoySocLondA)):

$$t_e(x, y) = -\frac{1}{2} i \varepsilon A^* \exp\left[-\frac{\pi}{4\mu}(\varepsilon - \sigma) r'^2\right] \exp\left(-\frac{\pi i r'^2}{2\mu}\right) \exp\left[-\frac{\pi \sigma}{2\mu}(r^2 + r_0^2)\right]. \tag{19}$$

In the following derivation, Gabor considers this twin (error) term for the case when $\varepsilon = \sigma$ (which is the same as $a = b$), which gives:

$$t_e(x, y) = -\frac{1}{2} i \sigma A^* \exp\left(-\frac{\pi i r'^2}{2\mu}\right) \exp\left[-\frac{\pi \sigma}{2\mu}(r^2 + r_0^2)\right]. \tag{20}$$

Next, Gabor represents the sample as a hexagonal (or close-packing) lattice of probability spots (as sketched in Fig. 7). This is done to be able to statistically average the signals in the reconstruction and quantitatively evaluate the contributions from the reconstructed object and twin (spurious) terms. For the lattice of probability spots, the reconstructed distribution is given by Eq. (17), where the first term describes a nearly constant background and the second (and third) terms are given by summation over all the reconstructed probability spots.

The spread of each probability spot is selected to be $a = b$. The distance between the adjacent probability spots is selected to be some distance $d$, which Gabor defines as the resolution limit: the probability spots can be resolved in the reconstruction if the minimum in the centre between the three closest probability spots just vanishes. By Eq. (17) the amplitude in the correct reproduction term follows $\exp\left[-\frac{\pi}{2}(r'/b)^2\right]$ if $a = b$. Thus, $d$ is found as follows: the three amplitudes from the closest reconstructed probability spots should add up to one at the coordinate between them. This gives an equation $\exp\left[-\frac{\pi}{2}\left(\frac{d}{b\sqrt{3}}\right)^2\right] = \frac{1}{3}$, and the solution is $d = 1.449 b$.

Each hexagonal cell around a probability spot occupies an area of

$$s_{hex} = d^2\sqrt{3}/2 = 0.866d^2 = 1.818b^2. \tag{21}$$

There is an error in Gabor's calculations by a factor of 2: $s_{hex}^{G} = d^2\sqrt{3}/4 = 0.433d^2 = 0.909b^2$.

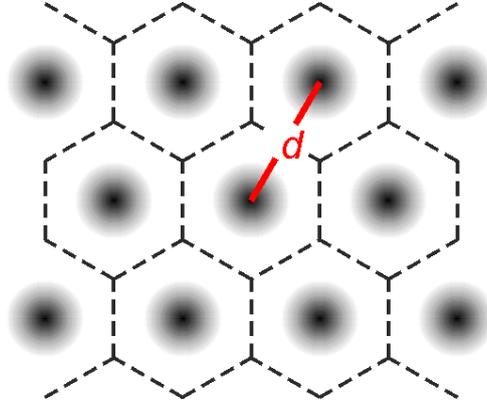

Fig. 7. Sample represented as a lattice of "probability spots".

Gabor defines the entire reconstructed region as an area of a radius $R$, where $R$ is the distance at which the background intensity drops by factor 1/3. To find $R$, one considers the reconstructed distribution given by Eq. (17), where the first term describes the background $t_r(x,y) \propto (1+i\sigma)\exp\left[-\frac{\pi\sigma}{\mu}(1+i\sigma)r^2\right] \approx \exp\left(-\frac{\pi\sigma}{\mu}r^2\right)$. $R$ is found by solving the equation $|t_r(x,y)|^2 \propto \exp\left(-\frac{2\pi\sigma}{\mu}R^2\right) = \frac{1}{3}$, which gives $R^2 \approx 0.175\frac{\mu}{\sigma}$. The number of the hexagonal cells (or, the hexagons) in the reconstructed area of radius $R$ is given by (Eq. (26) in ref. [2] (Gabor1949ProcRoySocLondA)):

$$N_{hexagons} = \frac{\pi R^2}{s_{hex}} = \frac{0.175\pi\mu}{1.818b^2\sigma} = \frac{0.302}{\sigma^2}, \tag{22}$$

where we used Eq. (18). $N_{hexagons}$ is the total number of hexagons. At the positions where no object is present (neither absorption nor phase shift occurs and the transmission is $t(x,y)=1$), the hexagon contains a probability spot with $A=0$.

Equation (20) describes the twin (or error, spurious) term in the reconstructed distribution of a single probability spot located at $(x_0, y_0)$. For a lattice of probability spots, Eq. (20) is modified by summing up the contributions from all the probability spots, where each probability spot is located at $(x_n, y_n)$ position (Eq. (27) in ref. [2] (Gabor1949ProcRoySocLondA)):

$$t_e(x,y) = -\frac{1}{2}i\sigma \exp\left(-\frac{\pi\sigma r^2}{2\mu}\right) \sum_{n=1}^{N_{\text{hexagons}}} A_n^* \exp\left(-\frac{\pi i r_n'^2}{2\mu}\right) \exp\left(-\frac{\pi\sigma r_n^2}{2\mu}\right), \qquad (23)$$

where $r_n^2 = x_n^2 + y_n^2$, and $r_n'^2 = (x-x_n)^2 + (y-y_n)^2$. By neglecting the slow-changing terms, this equation is approximated to (Eq. (27.1) in ref. [2] (Gabor1949ProcRoySocLondA))

$$t_e(x,y) \approx -\frac{1}{2}i\sigma \sum_{n=1}^{N_{\text{hexagons}}} A_n^* \exp\left(-\frac{\pi i r_n'^2}{2\mu}\right). \qquad (24)$$

This distribution provides the value of the twin term at a selected coordinate $(x,y)$ in the reconstructed distribution in the sample plane. The averaged intensity of this term evaluated as a statistical average is given by one-half of the sum of the absolute squares of the term on the right side:

$$t_{\text{eff}}^2 = \frac{1}{8}\sigma^2 \sum_{n=1}^{N_{\text{hexagons}}} A_n A_n^* = \frac{1}{8}\sigma^2 N_{\text{hexagons}} \overline{A_n A_n^*} = \frac{1}{8}\sigma^2 N_{\text{hexagons}} A_{\text{eff}}^2, \qquad (25)$$

where $\overline{A_n A_n^*}$ means averaging over the entire field, and

$$A_{\text{eff}} = \left(\overline{A_n A_n^*}\right)^{1/2} = \left(\frac{1}{N_{\text{hexagons}}} \sum_{n=1}^{N_{\text{hexagons}}} A_n A_n^*\right)^{1/2}. \qquad (26)$$

By substitution $N_{\text{hexagons}}\sigma^2 = 0.302$ from Eq. (22) into Eq. (25), one obtains:

$$t_{\text{eff}} = 0.194 A_{\text{eff}}. \qquad (27)$$

Equation (27) provides a very practical relation because it gives the averaged magnitude of the twin (or spurious, error) term at a selected coordinate in the reconstructed distribution ($t_{\text{eff}}$) being expressed directly through the parameters of the sample distribution ($A_{\text{eff}}$), without the need for simulation or reconstruction of hologram.

Gabor derived his criterion for objects suitable for holography based on the condition that the background in the reconstructed object's distribution should be nearly flat so that its intensity contrast does not exceed 0.05. For a signal with an average amplitude $U_{\text{avg}}$, the amplitude ranges from $U_{\min} = U_{\text{avg}} - \Delta U$ to $U_{\max} = U_{\text{avg}} + \Delta U$. The amplitude contrast is given by

$$C_{\text{ampl}} = \frac{U_{\max} - U_{\min}}{U_{\max} + U_{\min}} = \frac{\Delta U}{U_{\text{avg}}} \qquad (28)$$

and the intensity contrast is given by

$$C_{\text{int}} = \frac{I_{\max} - I_{\min}}{I_{\max} + I_{\min}} \approx \frac{2\Delta U}{U_{\text{avg}}}. \qquad (29)$$

According to Gabor, a background can be considered practically even if the intensity contrast $C_{int} \leq 0.05$, or the amplitude contrast $C_{ampl} \leq 0.025$. $t_{eff}$ is the deviation from the reconstructed background amplitude (assumed to be unity), which gives $C_{ampl} = \frac{\Delta U}{U_{avg}} = t_{eff}$, and therefore $C_{ampl} \leq 0.025$ means

$$t_{eff} \leq 0.025. \tag{30}$$

By using the relation given by Eq. (27), one obtains the condition when the reconstructed object is almost not affected by the twin (or spurious, error) signal:

$$A_{eff} \leq 0.129. \tag{31}$$

This condition is similar to the condition obtained by Gabor $A_{eff} < 0.1$.

### 4.1.1 Gabor's criterion for an opaque object

Gabor provides two examples: an amplitude-only opaque object and phase-only sample. Both samples are described by Eq. (8).

In the first example of an amplitude-only opaque object, it is assumed that a fraction $\kappa$ of the illuminated field is covered by an absorbing object where absorption occurs and $A = 1$. In the remaining fraction of the sample, no absorption occurs and $A = 0$. $A_{eff}$ is calculated by using Eq. (26):

$$A_{eff} = \left(\overline{A_n A_n^*}\right)^{1/2} = \left(\frac{1}{N_{hexagons}} \sum_{n=1}^{N_{hexagons}} A_n A_n^*\right)^{1/2} = \left(\frac{1}{N_{hexagons}} N_0\right)^{1/2} = \sqrt{\kappa}, \tag{32}$$

where $N_0$ is the number of hexagons with $A = 1$. Using Eq. (31), one obtains:

$$\kappa_{ampl} \leq 0.017. \tag{33}$$

This means that no more than about 2% (1% in ref. [2] (Gabor1949ProcRoySocLondA)) of the illuminated field should be covered by an opaque object (or dots, lines, etc.). The same criterion holds for a more general case of a non-opaque, amplitude-only object. For opaque and non-opaque objects of the same size, the twin term will be weaker in the case of the non-opaque object, and thus the conditions that the intensity contrast $C_{int} \leq 0.05$, or the amplitude contrast $C_{ampl} \leq 0.025$, will be certainly fulfilled and Eq. (33) can be used as a criterion in the general case of an amplitude-only object. The criterion that the object should occupy a finite region of the illuminated area is the most common criterion that is used when designing a holography experiment.

### 4.1.2 Gabor's criterion for phase-only object

The second example in Gabor's study is a sample with pure phase contrast with a random phase distribution that occupies the entire illuminated area. It is worth noting that such an object is different from the modern definition of a phase-shifting object, which is a finite-sized object with a non-random phase distribution.

In the case of the phase shifting object, Gabor assigns the average level of the transmission of the object to as a part of the background, and defines $A_{eff}$ as a measure of the departure from uniformity: $A_{eff}^2 = \overline{|A - \bar{A}|^2}$. In the case of random phase, the complex transmission vector $t = 1 - A$ moves on the unit circle, all orientations of $t$ are equally possible. This gives $\bar{t} = 0$, which in turn gives $\bar{A} = 1$, and $A_{eff}^2 = \overline{|A - \bar{A}|^2} = \overline{|A - 1|^2} = \overline{|t|^2} = 1$. The same result is obtained by considering the transmission function of a phase-only object $t(x, y) = 1 - A(x, y) = \exp[i\varphi(x, y)]$, which can be re-written as $A(x, y) = 1 - \exp[i\varphi(x, y)] = 1 - \cos[\varphi(x, y)] - i\sin[\varphi(x, y)]$ and which gives $\bar{A} = 1$ and $A_{eff}^2 = \overline{|A - \bar{A}|^2} = 1$ when $\varphi(x, y)$ is a random phase distribution. Thus, both approaches give $A_{eff}^2 = 1$. With this result of $A_{eff}^2 = 1$, Gabor concludes that the condition Eq. (31) cannot be fulfilled, and that a phase-shifting object cannot be reconstructed, even in principle. Although this derivation is correct, it considers an unpractical example of a sample with a random phase distribution that occupies the entire illuminated area. Below, we derive a criterion for a more practical case of a phase-shifting object that has a finite size.

### 4.1.3 Criterion for phase-only object

In this section, we derive a criterion for a phase-shifting object that occupies a finite-size region $\Omega$ and exhibits a random phase distribution by using an approach that is similar to Gabor's approach. The transmission function of a phase-shifting object can be written as:

$$t(x, y) = 1 - A(x, y) = \exp[i\varphi(x, y)], \tag{34}$$

which for a finite-size object gives:

$$t(x, y) = \begin{cases} \exp[i\varphi(x, y)], & (x, y) \in \Omega \\ 1, & (x, y) \notin \Omega \end{cases}, \tag{35}$$

and can be rewritten for $A(x, y)$

$$A(x,y) = \begin{cases} 1-\exp[i\varphi(x,y)], & (x,y) \in \Omega \\ 0, & (x,y) \notin \Omega \end{cases}. \tag{36}$$

This gives the average value of the transmission function

$$\bar{t} = \frac{\overline{\exp[i\varphi(x,y)]}N_0 + (N_{\text{hex}} - N_0)}{N_{\text{hex}}} = 1 - \kappa, \tag{37}$$

where $\overline{\exp[i\varphi(x,y)]} = 0$ is the average value over $N_0$ pixels; the object covers a fraction $\kappa$ of the illuminated field and is sampled with $N_0$ pixels. By substituting Eq. (37) into Eq. (34), we obtain $\bar{A} = 1 - \bar{t} = \kappa$. The deviation of $A(x,y)$ from its averaged value $\bar{A}$ is given by

$$A - \bar{A} = \begin{cases} 1-\exp[i\varphi(x,y)] - \kappa, & (x,y) \in \Omega \\ -\kappa, & (x,y) \notin \Omega \end{cases}, \tag{38}$$

which gives

$$|A - \bar{A}|^2 = \begin{cases} \left|1-\exp[i\varphi(x,y)] - \kappa\right|^2, & (x,y) \in \Omega \\ \kappa^2, & (x,y) \notin \Omega \end{cases},$$

and

$$A_{\text{eff}}^2 = \overline{|A - \bar{A}|^2} = \frac{\overline{\left|1-\exp[i\varphi(x,y)] - \kappa\right|^2}N_0 + \kappa^2(N_{\text{hex}} - N_0)}{N_{\text{hex}}} = 2\kappa - \kappa^2, \tag{39}$$

where $\overline{\left|1-\exp[i\varphi(x,y)] - \kappa\right|^2}$ is the average value over $N_0$ pixels. For small $\kappa$, such that $\kappa \ll 1$, one can approximate $A_{\text{eff}}^2 \approx 2\kappa$, and using the condition Eq. (31), one obtains:

$$\kappa_{\text{phase}} \leq 0.008. \tag{40}$$

A phase-shifting object can be reconstructed, provided it occupies less than 1% of the total illuminated area.

Experimental examples of in-line (Gabor) type holograms of phase-shifting can be found in numerous previous publications, for example, in refs. [6, 14-23] (Latychevskaia2007, Latychevskaia2009, Latychevskaia2010Ultramicroscopy, Jericho2012AO, Schwenke2012JMicroscopy, Rong2014OE, Rong2015SciRep, Wu2017JQuantSpectrRadTransfer, Lorenzo2018NanoLetters, Ling2021Microscopy, Latychevskaia2021PRL).

### 4.1.4 Simulated examples

To demonstrate how Gabor's criterion is fulfilled in practice, a hologram of a half-disk object that occupies slightly less than 0.8% of the imaged area is simulated. The object in the shape of a half-disk was also considered by Gabor in the example of an opaque object [2] (Gabor1949ProcRoySocLondA). In our study here, two cases were simulated: (1) an opaque object and (2) a phase-only object with a maximal shift of 1 radian. The parameters of the simulations were: the sample area size is 2 × 2 mm$^2$ sampled with 200 × 200 pixels, the source-to-sample distance is 40 mm, the source-to-detector distance is 0.1 m, the wavelength is 500 nm, and the object is a half-disk of 14 pixels in diameter (which makes it 0.73% of the imaged area). Although the simulations were done using spherical waves, the same results can be obtained with plane waves with the following parameters (as explained in ref. [10] (Latychevskaia2015AO)): sample area size of 2 × 2 mm$^2$, sample-to-detector distance of 40 mm, wavelength of 500 nm, sampled with 200 × 200 pixels. Simulation and reconstruction were performed using the algorithms described in ref. [10] (Latychevskaia2015AO). The artifacts due to the signal wrapping at the image edges during fast Fourier transforms were avoided by applying the following padding procedure: the sample transmission function of 200 × 200 pixels was padded up to 1000 × 1000 pixels. The calculated hologram of 1000 × 1000 pixels in size was cropped to 200 × 200 pixels to mimic the finite size of the detector. Then the cropped hologram was padded up to 1000 × 1000 pixels by constant background of one. The sample distribution reconstructed from the hologram was cropped to 200 × 200 pixels.

The obtained results are shown in Fig. 8, and these results demonstrate that for both types of objects: (i) the object can be qualitatively reconstructed; (ii) the reconstructed distributions are contaminated by the twin image; and, as a result, (iii) quantitatively, the amplitude and phase distributions are reconstructed with some error. Firstly, these results are already contradictory to Gabor's statement that, unlike an amplitude-only object, a phase object cannot be reconstructed from its hologram: objects of both types can be reconstructed. Secondly, even when Gabor's criterion is fulfilled, the amplitude-only object is not quantitatively correctly reconstructed. For both types of objects, the object distribution is reconstructed with some error.

The statistically averaged error term (defined by Eq. (25)) was calculated from the reconstructed amplitude and phase distributions using the following expressions:

$$\begin{aligned}\left(t_{\text{eff, ampl}}\right)^2 &= \frac{1}{2N^2} \sum_{i,j=0}^{N-1} \left|\text{rec}_e(i,j) - 1\right|^2 \\ \left(t_{\text{eff, phase}}\right)^2 &= \frac{1}{2N^2} \sum_{i,j=0}^{N} \left\{\text{Arg}\left[\text{rec}_e(i,j)\right]\right\}^2,\end{aligned} \quad (41)$$

where $\text{rec}_e(i,j)$ is the complex-valued distribution reconstructed from the background and twin image terms alone $(1+t_e)$ without the object term, and $i, j = 0...N-1$ are the pixel numbers. The calculated $t_{\text{eff}}$ are listed in Table 1. For both types of objects, the evaluated $t_{\text{eff}}$ is larger than required by the condition Eq. (30): $t_{\text{eff}} \leq 2.5\text{E-}2$. This can be explained by the fact that no masking function was applied to the hologram during the reconstruction, which can lead to larger values in the background and the twin image terms than those considered in the Gabor's derivation for a masked hologram.

The contrast of the signal due to the twin image term was evaluated as

$$C_{\text{ampl}} = \frac{\max|\text{rec}_e(i,j)| - \min|\text{rec}_e(i,j)|}{\max|\text{rec}_e(i,j)| + \min|\text{rec}_e(i,j)|}$$

$$C_{\text{phase}} = \left(\frac{1}{N^2}\sum_{i,j=0}^{N-1}\{\text{Arg}[\text{rec}_e(i,j)]\}^2\right)^{1/2}, \quad (42)$$

The contrast for the phase distribution was calculated using the root mean square expression because the "background" phase (or the average phase) is zero. The evaluated contrast values are summarized in Table 1. It should be noted that in the case of amplitude-only object, the contrast of the reconstructed signal due to the twin image term is 0.52, which is about ten times larger than the background intensity contrast value mentioned by Gabor.

To evaluate the mismatch between the original and reconstructed distributions, the mean squared error was calculated as

$$\text{MSE}_{\text{ampl}} = \frac{1}{N^2}\sum_{i,j=0}^{N-1}\left[|\text{rec}(i,j)| - |t(i,j)|\right]^2$$

$$\text{MSE}_{\text{phase}} = \frac{1}{N^2}\sum_{i,j=0}^{N-1}\{\text{Arg}[\text{rec}(i,j)] - \text{Arg}[t(i,j)]\}^2, \quad (43)$$

where $\text{rec}(i,j)$ is the reconstructed complex-valued distribution. The calculated MSE values are summarized in Table 1.

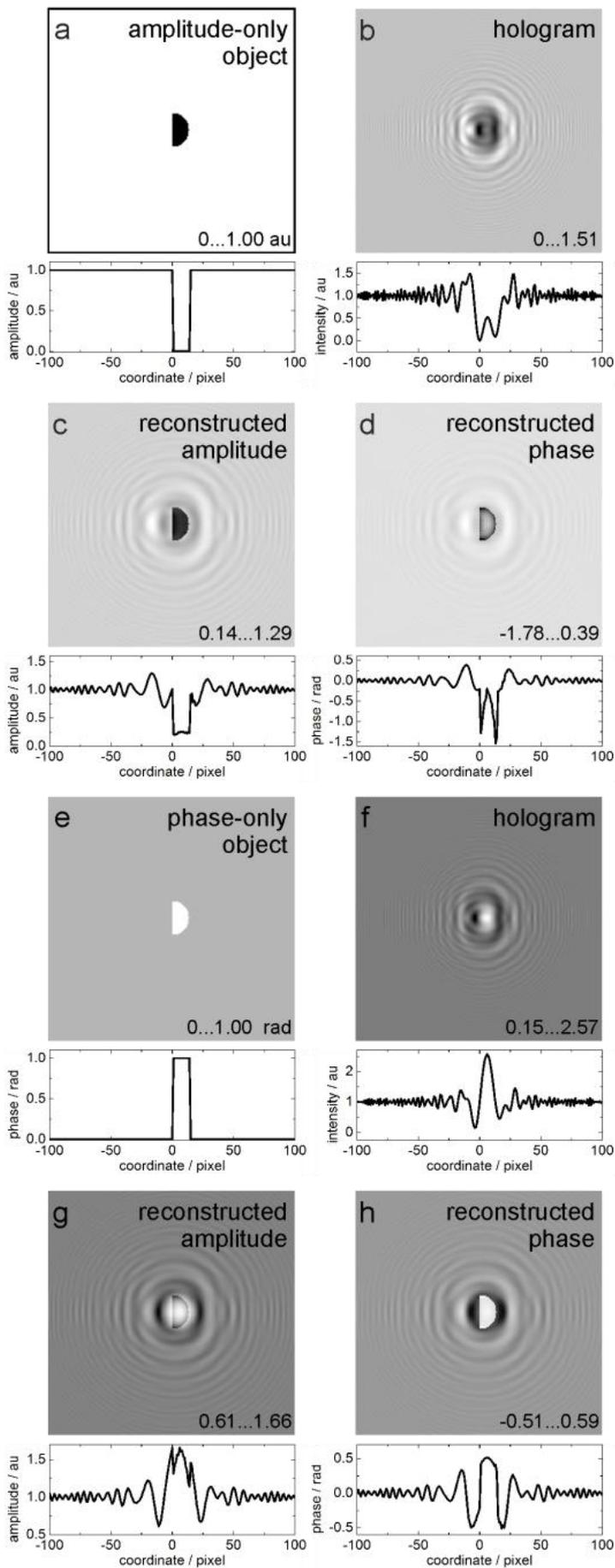

Fig. 8. Simulated examples of in-line (Gabor) type holograms of amplitude-only and phase-only objects in the form of a half-disk that occupies 0.73% of the imaged area. (a) – (d) Amplitude-only object: (a) The sample's amplitude distribution, (b) the corresponding hologram, (c) and (d) the reconstructed amplitude and phase distributions, respectively. (e) – (h) Phase-only object: (e) The sample phase distribution, (f) the corresponding hologram, (g) and (h) the reconstructed amplitude and phase distributions, respectively. The plots below the two-dimensional distributions show the horizontal profiles through the middle of the two-dimensional distributions.

| $\kappa$ / % | object type | $t_{\text{eff}}$ | contrast $C$ | MSE |
|---|---|---|---|---|
| 0.125 | amplitude-only | 1.675E-2 | 0.116 | 5.112E-4 |
|  | phase only | 1.661E-2 | 2.346E-2 | 5.476E-4 |
| 0.73 | amplitude-only | 4.056E-2 | 0.520 | 1.905E-3 |
|  | phase only | 4.5E-2 | 6.336E-2 | 5.038E-3 |
| 10 | amplitude-only | 0.164 | 0.994 | 7.824E-3 |
|  | phase only | 0.201 | 0.269 | 9.377E-2 |

Table 1. Errors of the reconstructed distributions of amplitude and phase-only objects for different fractions of the object to the imaged area $\kappa$. For phase-only objects $t_{\text{eff,phase}}$ and $C_{\text{phase}}$ are in radians, and $\text{MSE}_{\text{phase}}$ in radians$^2$.

## 4.2 Criteria derived using the Parseval's theorem

In this section, we provide a simplified derivation of the criteria for objects to be suitable for holography based on Parseval's theorem.

### 4.2.1 Amplitude-only object

In the case of an amplitude-only object, the transmission function of the sample can be represented as:

$$t(x, y) = 1 + o(x, y), \qquad (44)$$

and the total intensity of the object term is given by the sum

$$\Gamma = \sum_{i,j=1}^{N-1} |o(i, j)|^2. \qquad (45)$$

Both, the sample and hologram distributions are sampled with $N \times N = N^2$ pixels. The wavefront distribution on the detector plane is given by:

$$U_d(X, Y) = L[1 + o(x, y)] = R(X, Y) + O(X, Y), \qquad (46)$$

where $L$ is the operator of the wavefront forward propagation, and $L(1) = R$; $(X, Y)$ are the coordinates in the detector plane. In the holographic equation:

$$H = |U_d|^2 = |R|^2 + R^*O + RO^* + |O|^2 \qquad (47)$$

the object $R^*O$ and twin $RO^*$ terms carry the same energy (or intensity).

On the other hand, using the relation $L[o(x,y)] = O(X,Y)$ and the fact that the reference wave can be normalized to have amplitude of one, one can write Parseval's theorem for the total intensity of term $|o(x,y)|^2$ that should be equal to the total intensities of the $|R^*(X,Y)O(X,Y)|^2$ or $|R(X,Y)O^*(X,Y)|^2$ distributions:

$$\iint |o|^2 dxdy = \iint |O|^2 dXdY = \iint |R^*O|^2 dXdY = \iint |RO^*|^2 dXdY. \qquad (48)$$

Thus, knowing the sample distribution, one can find the total intensity of the twin signal from the total intensity of the object term by using Eq. (48).

The reconstructed distribution is given by:

$$L^{-1}(RH) \propto 1 + o + L^{-1}(R^2 O^*), \qquad (49)$$

where $L^{-1}$ is an operator of the wavefront backward propagation. In Eq. (49), the first two terms provide the exact reconstruction of the sample transmission function (as given by Eq. (44)). The third term in Eq. (49) describes the signal due to the twin term that is superimposed onto the reconstructed correct sample distribution. The total intensity of this twin term can be found by using Eq. (48).

In the reconstructed distribution, according to Eqs. (45), (48) and (49), the averaged intensity of the twin signal is $\overline{I_e} = \Gamma / N_{total}$, where $N_{total}$ is the total number of pixels ($N_{total} = N^2$). For an opaque object, $t(i,j) = 0$ at $N_0$ pixels, which gives $\Gamma = N_0$ and $\kappa = \dfrac{N_0}{N_{total}} = \dfrac{\Gamma}{N^2}$. Thus, the averaged intensity of the twin signal is $\overline{I_e} = \kappa$. The signal due to the twin image can be approximately described as $f\cos\left[\alpha(x^2 + y^2)\right]$, where $f$ and $\alpha$ are some constants. The average intensity of the twin image signal at a pixel is then given by $\overline{I_e} = \overline{f^2 \cos^2\left[\alpha(x^2 + y^2)\right]} = f^2/2$. Comparing both expressions for $\overline{I_e}$, one obtains $f = (2\kappa)^{1/2}$. In the reconstructed amplitude distribution of the background and twin image terms, the amplitude is varying from $A_{min} = 1 - f$ to $A_{max} = 1 + f$, which gives the contrast $C_{ampl} = \dfrac{A_{max} - A_{min}}{A_{max} + A_{min}} = f = \sqrt{2\kappa}.$

Using Gabor's definition of a uniform background, quote "a background can be considered as practically even if the intensity contrast does not exceed about 5%", for the reconstructed amplitude distribution, one obtains the following condition: $C_{\text{ampl}} = \sqrt{2\kappa} \leq 0.05$. This gives the criterion that the object should not occupy more than 0.125% of the illuminated area: $\kappa \leq 1.25\text{E}-3$ for the contrast due to the twin image term in the reconstructed amplitude distribution not to exceed 0.05.

Alternatively, one case use the requirement that the signal-to-noise ratio (SNR) of the reconstructed amplitude distribution should be at least 10, where the "signal" is the background of one and "noise" is the signal due to the twin image of amplitude $f$. This gives $\text{SNR} = 1/f \geq 10$, and taking into account that $f = (2\kappa)^{1/2}$, we obtain the following criterion for the object size

$$\kappa \leq 0.005. \tag{50}$$

To address the case of non-opaque amplitude objects, one can consider an opaque object that fulfils criterion Eq. (50) and a non-opaque object of the same size. For the non-opaque object, the twin term will be weaker than in the case of the opaque object, and therefore the criterion Eq. (50) will also be fulfilled. Thus, the criterion Eq. (50) can be used in the general case of an amplitude-only object.

### 4.2.2 Phase-only object

The transmission function of a phase-only sample can be written as:

$$t(x,y) = \exp[i\varphi(x,y)] \approx 1 + i\varphi(x,y). \tag{51}$$

The wavefront distribution on the detector plane is given by:

$$U_d(X,Y) = L[1 + i\varphi(x,y)] \approx R(X,Y) + i\Phi(X,Y), \tag{52}$$

where $L(\varphi) = \Phi$. The hologram distribution and the reconstructed sample distributions are given by

$$H \approx |R|^2 + iR^*\Phi - iR\Phi^* \tag{53}$$

and

$$L^{-1}(RH) \approx L^{-1}(R) + iL^{-1}(\Phi) - iL^{-1}(R^2\Phi^*) = 1 + i\varphi - iL^{-1}(R^2\Phi^*), \tag{54}$$

respectively. In Eq. (54), the first two terms are the correctly reconstructed signal, and the third term is the twin image term. The twin image term is an addition to the correctly reconstructed sample distribution. Neglecting the imaginary unit, Eq. (54) has the same appearance as in the case of the reconstructed amplitude-only object given by Eq. (49). The imaginary unit here plays the role of an indicator that the reconstructed distributions are associated with the phase distribution: $1+i\varphi-iL^{-1}\left(R^{2}\Phi^{*}\right)\approx\exp\left[i\left(\varphi+\Delta\varphi\right)\right]$ where $\Delta\varphi=-L^{-1}\left(R^{2}\Phi^{*}\right)$. Thus, in the case of a phase-only object, the amplitude (or intensity) of the twin image signal added to the correct phase distribution in the reconstruction (for example, as shown in Fig. 8h) can be analysed in the same way as the amplitude (or intensity) of the twin image signal added to the correct amplitude distribution in the reconstruction in the case of an amplitude-only object (for example, as shown in Fig. 8c). Thus, the same derivation as in the case of an amplitude-only object can be applied, and one obtains the same criterion that the phase-only object should not occupy an area larger than 0.5% of the illuminated area (Eq. (50)).

The calculations shown in Fig. 8 were also performed for a half-a-disk object that occupies 0.125% of the imaged area (the radius of the disk was 6 pixels). The simulations and reconstructions are not shown here, because qualitatively they look very similar to those shown in Fig. 8. $t_{\text{eff}}$, the contrast and MSE were calculated using Eqs. (41) – (43), respectively, and are summarized in Table 1. It should be noted that only for this object size, $t_{\text{eff}}$ and the contrast values are approaching 0.025 and 0.05, respectively – the values mentioned in Gabor's criterion of objects suitable for holography.

### 4.3 Iterative reconstruction in holography

Gabor derived the requirement for an object to be suitable for holography, assuming that: (1) reconstruction is made by simple wavefront backward propagation, (2) the signal due to the spurious (or error, twin image) term is distributed over the entire reconstructed area, and therefore its intensity at the object's position is much smaller than the intensity of the reconstructed correct term, (3) the signal due to the second-order object term $|O|^{2}$ is negligible.

When an object is reconstructed by applying an IPR algorithm, the requirements for the sample structure are the same as in CDI: the field of view should be larger than the object extent at least twice in each dimension. When an object is reconstructed by an IPR algorithm, the entire complex-valued field (the reference + object waves) $(R+O)$ is considered together as one complex-valued distribution, and as a result, the complex-valued transmission function is reconstructed $R+O\rightarrow r+o=t$. From the recovered transmission function $t$, the amplitude and phase distributions of the sample can be quantitatively correctly reconstructed.

For a hologram that is reconstructed by IPR methods, the requirement of the object size is much more generous than in the case of reconstruction by backward wavefront propagation. It means that for a hologram where the object does not fulfil Gabor's criterion but fulfils the IPR criterion, the object cannot be correctly reconstructed by using simple backward wavefront propagation, but it can be correctly reconstructed by using IPR methods. It should be mentioned though that conventional IPR methods are based on wavefront propagation back and forth between the sample and detector planes, and as a result, the reconstructed distribution is always a two-dimensional projection of the sample distribution.

As an example of an object that does not fulfil Gabor's criterion but fulfils the IPR criterion – a half-disk object that occupies 10% of the imaged area (the radius of the disk was 50.8 pixels) was studied, and the results are shown in Fig. 9. Here, qualitatively, the object can still be recognized in the reconstructed distributions obtained by simple wavefront backward propagation. But the correct amplitude or phase distributions cannot be extracted. $t_{\text{eff}}$, the contrast and MSE were calculated using Eqs. (41) – (43), respectively, and are summarized in Table 1. $t_{\text{eff}}$, the contrast and MSE exhibits relatively large values. In particular, it should be noted that the contrast of the signal due to the twin image is nearly one in the case amplitude-only object.

These results demonstrate that when an object is reconstructed from a Gabor (or in-line) hologram by non-iterative methods, the signal from the twin image term is always superimposed onto the reconstruction, and it increases as the ratio $\kappa$ increases. The amplitude and phase distributions of the object cannot be quantitatively correctly retrieved from such reconstructions. Gabor's criterion can be understood only as a requirement for the object distribution at which the background (due to the twin image and the second-order terms) in the reconstructed distribution to not exceed a certain level.

The results of iterative reconstruction of the object that occupies 10% of the imaged area from the same holograms are shown in Fig. 9i – j. The iteratively reconstructed amplitude and phase distributions exhibit no background signal and almost error-free recovered distributions. The mismatch between the original and iteratively reconstructed distributions (calculated by Eq. (41)) is $\text{MSE}_{\text{ampl}} = 1.458\text{E-4}$ and $\text{MSE}_{\text{phase}} = 2.155\text{E-4}$ rad$^2$ for amplitude-only and phase-only objects, respectively – which are very small values. The iterative reconstructions here were obtained by using the algorithms described previously in refs. [6, 7, 15] (Latychevskaia2007PRL, Latychevskaia2019JOSAA, Latychevskaia2010Ultramicroscopy) and applying 200 iterations. The constraints applied in the sample plane were: for the amplitude-only object – positive absorption and zero phase; for the phase-only object – positive phase, amplitude of one and a masking support that is larger by 2 pixels than the outer shape of the object (the signal outside the masking support was set to zero and it was not changed inside the masking support during the iterative routine) [15] (Latychevskaia2010Ultramicroscopy).

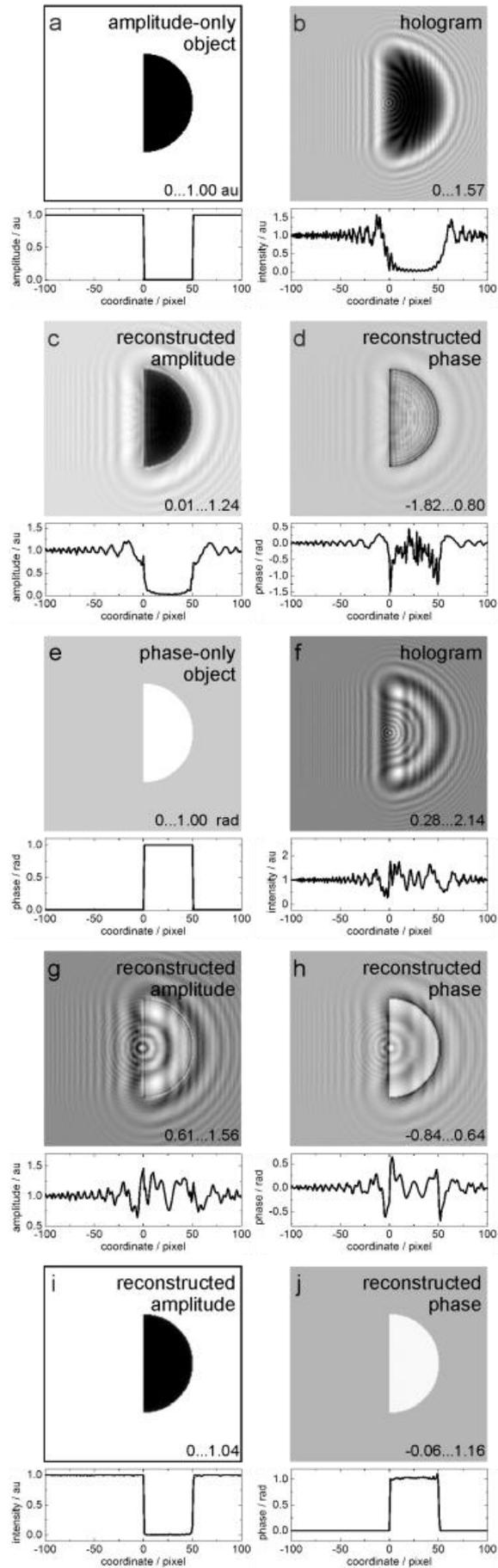

Fig. 9. Simulated examples of in-line (Gabor) type holograms of amplitude-only and phase-only objects in the form of a half-disk that occupies 10% of the imaged area (a) – (d) and (i) amplitude-only object: (a) The sample amplitude distribution, (b) the corresponding hologram, (c) and (d) the reconstructed amplitude and phase distributions, respectively. (e) – (h) and (j) phase-only object: (e) The sample phase distribution, (f) the hologram of the sample, (g) and (h) the reconstructed amplitude and phase distributions, respectively. (j) and (j) Amplitude and phase distributions obtained by iterative reconstruction of holograms in (b) and (f), respectively. The plots below the two-dimensional distributions show the horizontal profiles through the middle of the two-dimensional distributions.

## 5. Note on reconstruction of 3D objects

Although reconstruction of 3D objects from their digital holograms is outside of the scope of this study, one can add a short comment since holography is historically known as a technique that allows for the capture of 3D objects, where different parts of the object are located at different $z$-distances. Quasi-3D reconstruction of a 3D object can be obtained by either the simple wavefront back propagation method or by multi-slicing IPR methods [23] (Latychevskaia2021PRL). The criteria for the object being suitable for reconstruction in these cases refer to the size of the object's projection. The object's projection should occupy less than 1% of the illuminated area when the reconstruction is done by wavefront backpropagation. In the case where an IPR method is applied for reconstruction, the object's projection size should be less than half of the probed region in each dimension.

## 6. Discussion and conclusions

The Gabor criterion for objects suitable for holography is a condition at which the background in the reconstructed object's distribution can be considered nearly flat. The Gabor criterion is not related to the quantitative correctness of the reconstructed amplitude and phase distributions of the object (or sample). When an object occupies a larger area than provided by Gabor's criterion, the reconstruction by simple wavefront propagation can still be performed and the object can be recognized in the reconstruction. However, the reconstructed distribution will include all the terms, including the twin and second-order object terms. The signal due to these terms can be much smaller than that of the correct object distribution, but it is never zero, simply because the correct object's signal is not zero. As a result of these disturbances superimposed onto the reconstructed object's distribution, the amplitude (absorption) and phase distributions of the object cannot be quantitatively correctly retrieved.

The object's amplitude (absorption) and phase distributions can be quantitatively correct reconstructed only by applying IPR methods. To obtain a reconstruction by IPR algorithms, the parameters of the hologram acquisition must fulfil the same condition as in CDI: the imaged area should exceed at least twice the extent of the object in each direction. This condition is typically generously fulfilled in holography, where the objects are selected to fulfil Gabor's criteria and thus occupy only a small fraction of the imaged area that is much less than half of the extent of the imaged region. A small fraction, in turn, gives a large oversampling ratio. When applying IPR reconstruction methods, a larger oversampling ratio leads to a faster convergence of the IPR algorithm and a smaller error of the found solution.

## Appendix 1

The one-dimensional transformation from the sample plane to the Fourier plane (Eqs. (4) – (5)) of an exponential of quadratic function is given by the following pair (Eqs. (16.1) and (16.2) in ref. [2] (Gabor1949ProcRoySocLondA)):

$$t(x) = \exp\left[-\pi\left(A_0 x^2 + 2B_0 x\right)\right]$$

$$\tau(\varsigma) = \frac{1}{(1+i\mu A_0)^{1/2}} \exp\left[-\frac{\pi\mu}{1+i\mu A_0}\left(\mu A_0 \varsigma^2 + 2B_0 \varsigma - iB_0^2\right)\right].$$

## Appendix 2

The one-dimensional inverse transformation from the Fourier to the sample plane (Eqs. (4) – (5)) of an exponential of quadratic function is given by the following pair:

$$\tau(\varsigma) = \exp\left[-\pi\left(A_0 \varsigma^2 + 2B_0 \varsigma\right)\right].$$

$$t(x) = \frac{1}{(1+A_0/i\mu)^{1/2}} \exp\left[-\frac{\pi}{\mu(1+A_0/i\mu)}\left(\frac{A_0}{\mu}x^2 + 2B_0 x + iB_0^2\right)\right].$$

## Funding
Swiss National Science Foundation (SNSF) grant 197107.

## Acknowledgment
This work was funded by the Swiss National Science Foundation (SNSF) grant 197107.

## Disclosures
The authors declare no conflicts of interest.

## Data availability

Data underlying the results presented in this paper are not publicly available at this time but may be obtained from the authors upon reasonable request.